\documentclass[12pt]{article}
\usepackage{fullpage,hyperref,bbm,amssymb,amsfonts,amsthm}

\newcommand{\ket}[1]{|#1\rangle}
\newcommand{\bra}[1]{\langle #1|}
\newcommand{\<}{\langle}
\renewcommand{\>}{\rangle}

\newcommand{\F}{\mathbb{F}}
\newcommand{\Z}{\mathbb{F}}
\newtheoremstyle{dotless}{}{}{\itshape}{}{\bfseries}{}{ }
		{\thmname{#1}\thmnumber{ #2}\thmnote{ (#3)}:}
\theoremstyle{dotless}
\newtheorem*{definition}{Definition}
\newtheorem{lemma}{Lemma}
\newtheorem{theorem}{Theorem}
\newtheorem{corollary}{Corollary}

\date{}

\begin{document}
\title{Efficient Quantum Algorithm for Hidden\\ Quadratic and
Cubic Polynomial Function Graphs}
\author{Thomas Decker\thanks{Department of Computer Science \&
Engineering, University of Washington, Seattle, WA~98195, USA. 
Electronic address: \texttt{decker@ira.uka.de}}\, and
Pawel Wocjan\thanks{School of Electrical Engineering and Computer
Science, University of Central Florida, Orlando, FL~32816, USA. 
Electronic address: \texttt{wocjan@cs.ucf.edu}}}

\maketitle

\abstract{We introduce the Hidden Polynomial Function Graph Problem as
a natural generalization of an abelian Hidden Subgroup Problem (HSP)
where the subgroups and their cosets correspond to graphs of linear
functions over the finite field $\Z_p$. For the Hidden Polynomial
Function Graph Problem the functions are not restricted to be linear
but can also be multivariate polynomial functions of higher degree.

For a fixed number of indeterminates and bounded total degree the
Hidden Polynomial Function Graph Problem is hard on a classical
computer as its black box query complexity is polynomial in $p$.  In
contrast, this problem can be reduced to a quantum state
identification problem so that the resulting quantum query complexity
does not depend on $p$.  For univariate polynomials we construct a von
Neumann measurement for distinguishing the states.  We relate the
success probability and the implementation of this measurement to
certain classical problems involving polynomial equations.  We present
an efficient algorithm for hidden quadratic and cubic function graphs
by establishing that the success probability of the measurement is
lower bounded by a constant and that it can be implemented
efficiently.}

\section{Introduction}
Shor's algorithm for factoring integers and calculating discrete
logarithms \cite{Shor97} is one of the most important and well known
examples of quantum computational speedups. This algorithm as well as
other fast quantum algorithms for number-theoretic problems
\cite{Hallgren02,Hallgren05, SV05} essentially rely on the efficient
solution of an abelian hidden subgroup problem (HSP) \cite{BL95}.
This has naturally raised the questions of what interesting problems
can be reduced to the nonabelian HSP and of whether the general
nonabelian HSP can also be solved efficiently on a quantum computer.

It is known that an efficient quantum algorithm for the dihedral HSP
would give rise to efficient quantum algorithms for certain lattice
problems \cite{Regev02}, and that an efficient quantum algorithm for the
symmetric group would give rise to an efficient quantum algorithm for
the graph isomorphism problem \cite{EH99}. Despite the fact that
efficient algorithms have been developed for several nonabelian HSPs
(see, for example, \cite{ISS07} and the references therein), the HSP
over the dihedral group and the symmetric group have withstood all
attempts so far. Moreover, there is evidence that the nonabelian HSP
might be hard for some groups such as the symmetric group
\cite{Hallgren}.

Another idea for the generalization of the abelian HSP is to consider
Hidden Shift Problems \cite{BD07,DHI:03} or problems with hidden
non-linear structures \cite{Childs}. In the latter context, we define
a new black-box problem, called the Hidden Polynomial Function Graph
Problem, and present efficient quantum algorithms for special cases.
More specific, the Hidden Polynomial Function Graph Problem is a
natural generalization of the abelian HSP over groups of the special
form $G:=\F_p^{m+1}$, where the hidden subgroups are generated by the $m$
generators $(0, \ldots, 1 , \ldots, 0 ,q_i) \in \Z_p^{m+1}$ with
$q_i\in\Z_p$ and the $1$ is in the $i$th component. Therefore, the
hidden subgroups $H_Q$ and their cosets $H_{Q,z}$ are given by
\[
H_Q:=\{(x,Q(x)): x \in \Z_p^m \} \quad\mbox{and}\quad
H_{Q,z}:=\{(x,Q(x)+z): x \in \Z_p^m \}\,,
\]
where $z\in\Z_p$ and $Q$ runs over all polynomials 
$Q(X_1,\ldots,X_m)=q_1 X_1 + \ldots + q_m X_m$.  In the Hidden
Polynomial Function Graph Problem the polynomials
are no longer restricted to be linear but can also be of degree $n\ge
2$. The subgroups and their cosets are generalized to graphs of
polynomial multivariate functions going through the origin and to
translated function graphs, respectively.

Our approach to solve this problem on a quantum computer is to
generalize standard techniques for the HSP. First, we reduce the
problem to a quantum state identification problem and show that the
resulting quantum query complexity does not depend on $p$.  Second, we
design a measurement scheme for distinguishing the quantum states in
the univariate case. Third, we relate the success probability and
implementation of the measurement to certain classical problems
involving polynomial equations.

The paper is organized as follows: In Section~2 we define the Hidden
Polynomial Function Graph Problem and compare it to the Hidden
Polynomial Problem studied in Ref.~\cite{Childs}. In Section~3 we show
that the standard approach for HSPs can be used to reduce the new
problem to a state distinguishing problem. In Section~4 we derive
upper and lower bounds for the query complexity for this approach. In
Section~5 we discuss the properties of the states for univariate
polynomials and construct measurements to distinguish these states. In
Sections~6 and~7 we discuss the cases of quadratic and cubic
univariate functions thoroughly and show that an efficient solution
for these special cases exists. In Section~8 we conclude and discuss
possible objectives for further research.

\section{Hidden Polynomial Function Graph Problem}

\begin{definition}[Hidden Polynomial Function Graph Problem]${}$\\
Let $Q(X_1,X_2,\ldots,X_m)\in \Z_p[X_1,X_2,\ldots,X_m]$ be an
arbitrary $m$-variate polynomial of total degree at most $n$ whose constant
term is equal to zero.  Let $B : \Z_p^{m+1} \rightarrow \Z_p$ be a
black-box function hiding the polynomial $Q$ in the following sense:
\[
B(r_1,r_2,\ldots,r_m,s) = B(\bar{r}_1,\bar{r}_2,\ldots,\bar{r}_m,\bar{s})
\]
iff there is an element $z\in\Z_p$ such that
\[
s=Q(r_1,r_2,\ldots,r_m) + z \mbox{ and }
\bar{s}=Q(\bar{r}_1,\bar{r}_2,\ldots,\bar{r}_m) + z\,,
\]
i.e., the function $B$ is constant on the subsets 
\[
H_{Q,z} := \{(r_1,r_2,\ldots,r_m,Q(r_1,r_2,\ldots,r_m)+z)\,:\,
r_1,r_2,\ldots,r_m \in \Z_p \}
\] 
of $\Z_p^{m+1}$ and distinct for different values of $z$.

The Hidden Polynomial Function Graph Problem is to identify the
polynomial $Q$ if only the black-box function $B$ is given.  An
algorithm for $m$-variate polynomials with total degree less or equal
to $n$ (where $n$ and $m$ are both constant) is efficient if its
running time is polylogarithmic in $p$.
\end{definition}

An alternative definition of the function $B$ is
\[
B(r_1,r_2,\ldots,r_m,s):=\pi(s-Q(r_1,r_2,\ldots,r_m))
\]
where $\pi$ is an unknown and irrelevant bijection $\pi :
\Z_p\rightarrow \Z_p$ which permutes the elements of $\Z_p$
arbitrarily.

The classical query complexity of the Hidden Polynomial Function Graph Problem is polynomial in $p$.  This
is because for univariate polynomials (i.e., $m=1$) at least $n$
different points
\[
(r^{(1)},s^{(1)}),\ldots,(r^{(n)},s^{(n)}) \quad {\rm  with} \quad 
B(r^{(1)},s^{(1)})=\ldots=B(r^{(n)},s^{(n)})
\]
are required in order to determine the hidden polynomial $Q$ of degree
$n$.  The probability of obtaining such an $n$-fold collision is smaller
than the probability of obtaining a $2$-fold collision.  The probability of
the latter is $1/p$.  

The Hidden Polynomial Function Graph Problem is related to the Hidden
Polynomial Problem defined in \cite{Childs} which can be equivalently
reformulated as follows. The black-box function $h : \Z_p^m
\rightarrow \Z_p$ is given by $h(r_1, \ldots, r_m):=\sigma(Q(r_1,
\ldots, r_m))$, where $\sigma$ is an arbitrary permutation of $\Z_p$
and $Q(X_1, \ldots, X_m)$ is the hidden polynomial.  It is readily
seen that the black-boxes $h$ can be obtained from the black-boxes $B$
by querying $B$ only at points of the form $(r_1,\ldots,r_m,0)$.  For
this reason the black-boxes $B$ offer more flexibility in designing
quantum algorithms.  We are able to design an efficient quantum
algorithm for the black-boxes $B$ hiding univariate quadratic and
cubic polynomials, whereas no algorithms are known for the black-boxes
$h$.

\section{Standard Approach}
Most quantum algorithms for HSPs are based on the standard approach
which reduces black box problems to state distinguishing problems.  We
apply this approach to the Hidden Polynomial Function Graph Problem in
the following.

\begin{itemize}
\item Evaluate the black-box function on an equally weighted
superposition of all $(r_1,r_2,\ldots,r_m,s) \in \Z_p^{m+1}$.  The
resulting state is
\[
\frac{1}{\sqrt{p^{m+1}}} \sum_{r_1,r_2,\ldots,r_m,s\in \Z_p} \ket{r_1,
r_2, \ldots, r_m} \otimes \ket{s} \otimes \ket{F(r_1,r_2,\ldots,r_m,s)}
\]
\item Measure and discard the third register.  Assume we have obtained 
the result $\pi(z)$.  Then the state 
on the first and second register is $\rho_{Q,z}:= \ket{\phi_{Q,z}} 
\bra{\phi_{Q,z}}$ where 
\[ 
|\phi_{Q,z}\> := \frac{1}{\sqrt{p^m}} \sum_{r_1,r_2,\ldots,r_m\in
  \Z_p} \ket{r_1, r_2, \ldots, r_m} \otimes
\ket{Q(r_1,r_2,\ldots,r_m,s)+z}
\]
with the unknown polynomial $Q$ hidden by $B$, and $z$ is uniformly
at random.  The corresponding density matrix is
\begin{equation}\label{EQ EQ}
\rho_Q := \frac{1}{p} \sum_{z\in\Z_p} |\phi_{Q,z}\> \<\phi_{Q,z}|\,.
\end{equation}
\end{itemize}

We refer to the states $\rho_Q$ as {\em polynomial function
states}. We have to distinguish these states in order to solve the black box
problem.

\section{Quantum Query Complexity}

We show that the quantum query complexity of the Hidden Polynomial
Function Graph Problem is independent of $p$.  To prove this result we
make use of the upper and lower bounds of Ref.~\cite{HW06} on the
number of copies required for state discrimination.  The former is
expressed in terms of fidelity which can be bounded by the following
technical lemma.

\begin{lemma}
Let $\rho$ and $\sigma$ be two quantum states with corresponding
spectral decompositions $\rho=\sum_i \lambda_i |\psi_i\>\<\psi_i|$ and
$\sigma = \sum_j \mu_j |\phi_j\>\<\phi_j|$.  Assume that
$\max_{i,j}|\<\psi_i|\phi_j\>|\le\alpha$ for some value $\alpha$.
Then we have
\[
F(\rho,\sigma)\le \alpha  \cdot 
\min
\left\{\;
\sum_i \sqrt{\lambda_i}\,, \;
\sum_j \sqrt{\mu_j} \;
\right\}
\,,
\]
where $F(\rho,\sigma):=\|\sqrt{\rho} \sqrt{\sigma}\|_1$ is the
fidelity of $\rho$ and $\sigma$.
\end{lemma}

\begin{proof}
We have 
\begin{equation}\label{eq:trick}
\|\sqrt{\rho}|\phi_i\>\<\phi_i|\|_1 \le \alpha
\end{equation}
for all $i$.  This is derived by observing that
$\|\sqrt{\rho}|\phi_i\>\<\phi_i|\|_1 = \|\sqrt{\rho}|\phi_i\>\| \,
\||\phi_i\>\|$ and
\[
\|\sqrt{\rho}|\phi_j\>\|^2\le\sum_i\lambda_i |\<\psi_i|\phi_j\>|^2 \le
\alpha^2\,.
\]
Using first the triangle inequality and then Eq.~(\ref{eq:trick}) we
obtain
\begin{equation}
\| \sqrt{\rho} \sqrt{\sigma} \|_1 = \| \sqrt{\rho} \sum_j \sqrt{\mu_j}
|\phi_j\>\<\phi_j| \|_1 \le \sum_j \sqrt{\mu_j} \| \sqrt{\rho}
|\phi_j\>\<\phi_j| \|_1 = \alpha \cdot \left( \sum_j \sqrt{\mu_j}
\right)\,.
\end{equation}
The same arguments apply if we use the spectral decomposition of
$\rho$ instead.  This completes the proof.
\end{proof}

\begin{corollary}
We have $F(\rho_Q,\rho_{\tilde{Q}})\le n/\sqrt{p}$, where $\rho_Q$ and
$\rho_{\tilde{Q}}$ are two different polynomial states and their total
degree is at most $n$.
\end{corollary}
This corollary follows by observing that 
\begin{eqnarray*}
|\<\phi_{Q,z}|\phi_{\tilde{Q},\tilde{z}}\>| & = &
\frac{1}{p^m}\sum_{r_1,\ldots,r_m\in\Z_p} \<Q(r_1,\ldots, r_m)+z |
\tilde{Q}(r_1,\ldots, r_m)+ \tilde{z}\> \\ & = & \frac{1}{p^m}|\{
(r_1,\ldots,r_m)\in\Z_p^m\, :\, Q(r_1,\ldots,r_m) + z =
\tilde{Q}(r_1,\ldots,r_m) + \tilde{z}\}| \\ & \le & \frac{1}{p^m} \, n
\, p^{m-1} = \frac{n}{p}\,.
\end{eqnarray*}
The last inequality follows from the Schwartz-Zippel theorem saying
that two different $m$-variate polynomials of total degree less or
equal to $n$ can intersect in at most $n p^{m-1}$ points \cite{MR95}.

\begin{theorem}
The query complexity of the Hidden Function Graph Problem is at most
$4 {n+m\choose m}$.
\end{theorem}
\begin{proof}
The results in \cite{HW06} imply that there is a POVM $\{E_Q\}$ acting
on $k$ copies of a polynomial function state such that
\[
P_{\mathrm{success}}:= \min_Q
\mathrm{Tr}(\rho_Q^{\otimes k} E_Q) \ge 1 - \epsilon
\]
provided that $k \ge 2(\log N - \log \epsilon)/(-\log F)$, where
$N:=p^{{n+m \choose n}-1}$ is the number of different polynomial
function states and $F$ is the maximal fidelity over all pairs of
different polynomial function states.  This bound and the lower bound
on the fidelity $F\le n/\sqrt{p}$ imply that the success probability
$P_{\mathrm{success}}$ is at least $1/2$ for $k=4 {n+m \choose n}$
(provided that $p$ is sufficiently large).
\end{proof}

The lower bound presented in \cite{HW06} implies that at least ${n+m
\choose m}/m-1$ copies are required to have $P_{\mathrm{success}}\ge
1/2$.

\section{Distinguishing Polynomial Function States}
In the remainder of the article we consider only the univariate case,
i.e., $m=1$.

\medskip 
\noindent
{\bf Structure of Polynomial Function States}\quad The states
$\rho_{Q,z}$ can be written as
\[
\rho_{Q,z}= \frac{1}{p}\sum_{b,c\in\Z_p} \ket{b}\bra{c} \otimes
\ket{Q(b)+z}\bra{Q(c)+z}\,.
\]
The density matrix $\rho_Q$ of Eq.~(\ref{EQ EQ}) is the average of
these states over $z$. To obtain a compact notation we introduce the
cyclic shift $ S_p|x\>:=\ket{x+1 \; {\rm mod} \; p}$ for which we have
the identity
\[
\sum_{z \in \Z_p} \ket{b+z}\bra{c+z} = S_p^{b-c}\,.
\]
This directly leads to 
\[
\rho_Q= \frac{1}{p^2}
\sum_{b,c\in\Z_p} \ket{b}\bra{c} \otimes S_p^{Q(b)-Q(c)}\,.
\]

Now we use the fact that the shift operator and its powers can
be diagonalized simultaneously with the Fourier matrix
$
F_p:= \sqrt{1/p} \sum_{k,\ell\in\Z_p} \omega_p^{k \ell}
\ket{k}\bra{\ell}
$,
i.e., we have 
\[
F_p S_p^k F_p^\dagger = \sum_{u \in \Z_p} \omega_p^{uk}
\ket{u}\bra{u}\,,
\]
where $\omega_p:=e^{2\pi i/p}$ is a $p$th root of unity. Hence, the 
density matrices have the block diagonal form
\[
\tilde{\rho}_Q := (I_p \otimes F_p)\rho_Q (I_p \otimes F_p^\dagger)= 
\frac{1}{p^2} \sum_{b,c,x\in\Z_p}
\omega_p^{[Q(b)-Q(c)]x} |b\>\<c| \otimes |x\>\<x|
\]
in the Fourier basis where $I_p$ denotes the identity matrix of size
$p$.

By repeating the standard approach $k$ times for the same black-box
function $B$, we obtain the density matrix $\tilde{\rho}_Q^{\otimes
k}$. After rearranging the registers we can write
\begin{eqnarray*}
\tilde{\rho}_Q^{\otimes k} & = & \frac{1}{p^{2k}}
\sum_{b,c,x\in\Z_p^k} \omega_p^{\sum_{j=1}^k [Q(b_j)-Q(c_j)] x_j}
|b\>\<c| \otimes |x\>\<x| \\
& = & \frac{1}{p^{2k}} \sum_{b,c,x\in\Z_p^k} \omega_p^{\sum_{j=1}^k
[\sum_{i=1}^n q_i(b_j^i - c_j^i)]x_j} |b\>\<c| \otimes |x\>\<x| \\
& = & \frac{1}{p^{2k}} \sum_{b,c,x\in\Z_p^k} \omega_p^{\sum_{i=1}^n
q_i [\sum_{j=1}^k (b_j^i - c_j^i)x_j]} |b\>\<c| \otimes |x\>\<x| \\
& = & \frac{1}{p^{2k}} \sum_{b,c,x\in\Z_p^k}
\omega_p^{\<q|\Phi^{(n)}(b) - \Phi^{(n)}(c)|x\>} |b\>\<c| \otimes
|x\>\<x|\,,
\end{eqnarray*}
where $\<q|$, $\Phi^{(n)}(b)$, $\Phi^{(n)}(c)$, and $|x\rangle$ are
defined as follows:
\begin{itemize}
\item $\<q|:=(q_1,q_2,\ldots,q_n) \in \Z_p^{1 \times n} $ is the row
vector whose entries are the coefficients of the hidden polynomial
$Q(X)=\sum_{i=1}^n q_i X^i$, 
\item $\Phi^{(n)}(b)$ is the $n\times k$ matrix
\[
\Phi^{(n)}(b) := \sum_{i=1}^n \sum_{j=1}^k b_j^i |i\>\<j| = \left(
\begin{array}{cccc}
b_1    & b_2    & \cdots & b_k    \\
b_1^2  & b_2^2  & \cdots & b_k^2  \\
\vdots & \vdots &        & \vdots \\
b_1^n  & b_2^n  & \cdots & b_k^n  
\end{array}
\right)\,,
\]
\item
$|x\>:=(x_1,\ldots,x_k)^T \in \Z_p^k$ is the column vector whose
entries are those of $x$.
\end{itemize}

\noindent
{\bf Algebraic-geometric problem}\quad We simplify the techniques of
\cite{BCD06,BCD05,BD07} and use them to construct a von Neumann
measurement for distinguishing the states $\tilde{\rho}_Q^{\otimes
k}$.  Let $w:=(w_1,\ldots,w_n)\in\Z_p^n$ and $|w\>\in\Z_p^{n}$ be the
corresponding column vector.  Consider the algebraic-geometric problem
to determine all $b \in \Z_p^k$ for given $x\in \Z_p^k $ and $w \in
\Z_p^n$ such that $\Phi^{(n)}(b)|x\> = |w\>$, i.e.,
\[
\left(
\begin{array}{cccc}
b_1 & b_2 & \cdots & b_k \\ b_1^2 & b_2^2 & \cdots & b_k^2 \\ \vdots &
\vdots & & \vdots \\ b_1^n & b_2^n & \cdots & b_k^n
\end{array}
\right)\cdot \left(
\begin{array}{c}
x_1 \\ x_2 \\ \vdots \\ x_k
\end{array}
\right) = \left(
\begin{array}{c}
w_1 \\ w_2 \\ \vdots \\ w_n
\end{array}
\right)
\]
We denote the set of solutions to these polynomial equations and its
cardinality by
\[
S_w^x := \{ b\in \Z_p^k \,:\, \Phi^{(n)}(b)|x\> = |w\>\} 
\quad\; {\rm and} \quad\;
\eta_w^x := |S_w^x|\,,
\]
respectively. We also define the quantum states $|S_w^x\>$ to be the
equally weighted superposition of all solutions
\[
|S_w^x\> := \frac{1}{\sqrt{\eta_w^x}} \sum_{b\in S_w^x} |b\>
\]
if $\eta_w^x >0$ and $\ket{S_w^x}$ to be the zero vector otherwise.
Using this notation, we can express the state $\tilde{\rho}_Q^{\otimes
k}$ as
\[
\tilde{\rho}_Q^{\otimes k} := \frac{1}{p^{2k}}\sum_{x\in\Z_p^k}
\sum_{w,v\in\Z_p^n} \omega_p^{\<q|w\> - \<q|v\>} \sqrt{\eta_w^x
\eta_v^x} |S_w^x\>\<S_v^x| \otimes |x\>\<x|\,.
\]

\noindent {\bf Measurement for distinguishing the polynomial
states}\quad
The block structure of the states $\tilde{\rho}_Q^{\otimes k}$ implies
that we can measure the second register in the computational basis
without any loss of information.  The probability of obtaining a
particular $x$ is
\[
\mathrm{Tr}\left(\tilde{\rho}_Q^{\otimes k} (I_{p^k}\otimes
|x\>\<x|)\right) = \frac{1}{p^{2k}} \sum_{w\in\F_p^n} \eta_w^x =
\frac{1}{p^k}
\]
and the resulting reduced state is
\begin{equation}\label{eqeq3}
\tilde{\rho}_Q^x := \frac{1}{p^k} \sum_{w,v\in\Z_p^n}
\omega_p^{\<q|w\> - \<q|v\>} \sqrt{\eta_w^x \eta_v^x}
|S_w^x\>\<S_v^x|\,.
\end{equation}
In the following we assume that for a result $x$ and all $w$ the
cardinality $\eta_w^x$ is at most polylogarithmic in $p$ and that the
elements of the sets $S^x_w$ can be computed efficiently.  In this
case we have an efficiently computable bijection between $S_w^x$ and
the set $\{ (w,j) \,:\, j=\{0,\ldots,\eta_w^x-1\} \}$.  This bijection
is obtained by sorting the elements of $S_w^x$ according to the
lexicographic order on $\F_p^k$ and associating to each $b\in S_w^x$
the unique $j\in\{0,\ldots,\eta_w^x-1\}$ corresponding to its position
in $S_w^x$.  We rely on this bijection to implement a transformation
$U_x$ satisfying
\[
U_x |S_w^x\> = |w\>
\]
for all $(x,w)$ with $\eta_w^x>0$.  This is done as follows.
\begin{itemize}\label{eQ5}
\item Implement a unitary with
\begin{equation}
\frac{1}{\sqrt{\eta_w^x}} \sum_{b \in S_w^x} \ket{b} 
\otimes \ket{0} \otimes \ket{0} \mapsto 
\frac{1}{\sqrt{\eta_w^x}} \ket{w} \otimes \sum_{j=1}^{\eta^x_w} 
\ket{j} \otimes \ket{\eta_w^x}\,.
\end{equation}
Note that $b$ and $x$ determine $j$ and $w$ uniquely and vice versa.
Furthermore, we can compute $w$ and $j$ efficiently since $\eta_w^x$
is at most polylogarithmic in $p$.

\item Apply the unitary 
\[
\sum_{\ell=0}^{\eta_w^x-1} (F_{\ell+1} \oplus I_{p^k-\ell-1}) \otimes
\ket{\ell}\bra{\ell} + \sum_{\ell=\eta_w^x}^{p^k-1} 
I_{p^k} \otimes \ket{\ell}\bra{\ell}
\]
on the second and third register. This implements the embedded Fourier
transform $F_\ell$ of size $\ell$ controlled by the second register in
order to map the superposition of all $\ket{j}$ with $j \in \{0,
\ldots, \ell-1\}$ to $\ket{0}$. The resulting state is $\ket{w}
\otimes \ket{0} \otimes \ket{\eta_w^x}$.

\item Uncompute $\ket{\eta_w^x}$ in the third register with the help
of $w$ and $x$. This leads to the state $\ket{w}\otimes \ket{0}
\otimes \ket{0}$
\end{itemize}

We apply $U_x$ to the state of Eq.~(\ref{eqeq3}) and obtain
\[
U_x \tilde{\rho}_Q^x U_x^\dagger = \frac{1}{p^k} \sum_{w,v\in\Z_p^n}
\omega_p^{\<q|w\> - \<q|v\>} \sqrt{\eta_w^x \eta_v^x} |w\>\<v|\,.
\]

We now measure in the Fourier basis, i.e., we carry out the von
Neumann measurement with respect to the states
\[
|\psi_Q\> := \frac{1}{\sqrt{p^n}} \sum_{w \in \F_p^n}
 \omega_p^{\<q|w\>} |w\>\,.
\]
Simple computations show that the probability for the correct
detection of the state ${\tilde \rho}_Q^x$ is
\begin{equation}\label{eq:probForSpecific_x}
\langle \psi_Q | {\tilde \rho}_Q^x | \psi_Q \rangle =
\frac{1}{p^{k+n}} \left( \sum_{w \in \F_p^n} \sqrt{\eta_w^x}
\right)^2\,.
\end{equation}
The probability to identify $Q$ correctly is obtained by summing the
probabilities in Eq.~(\ref{eq:probForSpecific_x}) over all $x$ for
which we can implement the transformation $U_x$ and multiplying the
sum by $1/p^k$.

\section{Hidden Quadratic Polynomials}\label{sec 6}
For a single copy of the polynomial function state $\rho_Q$ it turns
out that the pretty good measurement~\cite{HW94} is the optimal
measurement for distinguishing the states. However, the resulting
success probability is only in the order of $1/p$. In contrast, the
success probability of our measurement scheme for two copies is lower
bounded by a constant. This strongly resembles the situation for the
Heisenberg-Weyl HSP, where a single copy is also not sufficient but
the pretty good measurement of two copies leads to an efficient
quantum algorithm \cite{BCD05}.

For quadratic polynomials we have to consider the sets
\[
S_{(w_1,w_2)}^{(x_1,x_2)}= \left\{ 
\left( \begin{array}{c}b_1\\b_2\end{array}\right) \in \Z_p^2 \;\;:\;\;
\left( \begin{array}{cc}b_1&b_2\\b_1^2&b_2^2\end{array}\right) 
\cdot \left(
\begin{array}{c}x_1\\x_2 \end{array}\right) =
\left(\begin{array}{c}w_1\\w_2 \end{array}\right)\right\}\,.
\]
We set $b=b_1$, $c=b_2$, $x=x_1$, $y=x_2$, $v=w_1$, and $w=w_2$ to
avoid too many indices. Therefore, we have to find the set of
solutions of the equations
\begin{equation}\label{EQ ZWEI}
bx+cy=v \quad {\rm and} \quad b^2x+c^2y=w\,.
\end{equation}
Depending on $x$ and $y$ which are determined by the orthogonal
measurement in the first stage as well as by $v$ and $w$ the set of
solutions can encompass $0$, $1$, $2$, $p$ or $p^2$ solutions.  To
derive a lower bound on the success probability 
it suffices to consider the $p^2-3p+2$ cases where $x,y \not=0$ and
$x\not= -y$.  In these cases the Eqs.~(\ref{EQ ZWEI}) have the
solutions $(b_j,c_j)$ with
\[
c_{1/2}:= \frac{v}{x+y} \pm \frac{1}{x+y}\sqrt{D} \quad \mbox{and}
\quad b_{1/2}=\frac{v}{x}-\frac{y}{x}c_{1/2}
\]
provided that
\[
D:=\frac{x}{y} w ((x+y)- v^2)
\]
is a square in $\Z_p$.  For each pair $(x,y)$ there are $p(p+1)/2$
pairs $(v,w)$ such that the resulting $D$ is a square.  In this case,
there are one or two solutions. Therefore, we have the following lower
bound on the success probability
\[
\frac{1}{p^6} \sum_{(x,y)} \left( \sum_{(v,w)}
\sqrt{\eta_{(v,w)}^{(x,y)}} \right)^2 \geq \frac{1}{p^6}\,
(p^2-3p+2)\, \left(\frac{p(p+1)}{2}\right)^2=
\frac{1}{4}-O\left(\frac{1}{p}\right)\,.
\]
We now argue that the measurement can be implemented
efficiently. Following the discussion of Sec.~5 we only have to show
that we can implement the transform of~(\ref{eQ5}) efficiently, i.e.,
given $x,y,b$ and $c$ we must find the index of the solution $(b,c)$
to Eqs.~(\ref{EQ ZWEI}) efficiently. This is possible since the
solutions of the $p^2-3p+2$ considered cases can be computed with
$O({\rm log}(p))$ operations on a classical computer (see Cor.~14.16
in~\cite{Gathen}).

\section{Hidden Cubic Polynomials}\label{sec7}
For cubic polynomials we obtain the sets
\begin{equation}\label{Problem3}
S_{(w_1,w_2,w_3)}^{(x_1,x_2,x_3)} = \left\{ \left(
\begin{array}{c}b_1\\b_2\\b_3\end{array}\right) \in \F_p^3 : \left(
\begin{array}{ccc} b_1 & b_2 & b_3 \\ b_1^2 & b_2^2 & b_3^2 \\ b_1^3 &
b_2^3 & b_3^3 \end{array} \right) \cdot \left( \begin{array}{c}x_1 \\
x_2 \\ x_3 \end{array} \right)= \left( \begin{array}{c}w_1 \\ w_2 \\
w_3 \end{array} \right) \right\}\,.
\end{equation}
To simplify the following computations we assume that $x_1 \not
=0$. Therefore, the set of Eq.~(\ref{Problem3}) can be written as
\begin{equation}\label{EQQ}
S^\kappa_\lambda=\left\{
\left(\begin{array}{c}b\\c\\d\end{array}\right)
\in \F_p^3 : \left( \begin{array}{ccc} b & c & d \\
b^2 & c^2 & d^2 \\ b^3 & c^3 & d^3 \end{array} \right)
\cdot \left( \begin{array}{c}1 \\ x \\ y \end{array} \right)=
\left( \begin{array}{c}u \\ v \\ w \end{array} \right) \right\}
\end{equation}
with $\kappa:=(1,x,y)$, $\lambda:=(u,v,w)$, and the coefficients
\[
x:=\frac{x_2}{x_1}\,,\;\;\;
y:=\frac{x_3}{x_1}\,,\;\;\;
u:=\frac{w_1}{x_1}\,,\;\;\;
v:=\frac{w_2}{x_1}\,,\;\;\;{\rm and}\;\;\;
w:=\frac{w_3}{x_1}\,.
\]
In the appendix we show that for
\begin{equation}\label{Ineq}
x\not=0,\pm 1\;\;\; {\rm and} \;\;\;
y\not=0,-1,-x,\pm(x+1)
\end{equation}
and for all $u,v,w$ the inequality
\begin{equation}\label{ineq2}
\eta^\kappa_\lambda \leq 10
\end{equation}
holds for the size $\eta_\lambda^\kappa$ of the sets of
Eq.~(\ref{EQQ}). This bound now implies that for all pairs $(x,y)$
there are at least $p^3/10$ tuples $(u,v,w)$ with $\eta^\kappa_\lambda\ge
1$ because of the equality
\[
\sum_{\lambda \in\F_p^3}\eta_\lambda^\kappa=p^3\,.
\]

We obtain a lower bound on the success probability $P_{\rm success}$ for our
measurement scheme as follows. First, we discard all tuples
$(x_1,x_2,x_3,w_1,w_2,w_3)$ with $x_1=0$. This leads to
\begin{eqnarray*}
P_{\rm success}&=& 
\frac{1}{p^9}\sum_{x_1,x_2,x_3 \in \F_p} \left( \sum_{w_1,w_2,w_3\in\F_p} 
\sqrt{\eta_{(w_1,w_2,w_3)}^{(x_1,x_2,x_3)}}\right)^2\\
& \geq & \frac{p-1}{p^9}\sum_{x,y\in \F_p} \left( 
\sum_{u,v,w \in \F_p}  
\sqrt{\eta_{(u,v,w)}^{(1,x,y)}}\right)^2\,.
\end{eqnarray*}
Second, we take the disequalities~(\ref{Ineq})
into account and obtain 
\[
P_{\rm success} \geq \frac{(p-1)(p^2-8p+16)}{p^6} 
\left( \sum_{u,v,w \in \F_p}
\sqrt{\eta^{(1,x,y)}_{(u,v,w)}}\right)^2
\]
because there are $(p-3)(p-5)+1$ pairs $(x,y)$ which satisfy these
disequalities. Third, we lower bound the sum by $p^3/10$ 
and obtain
\[
P_{\rm success} \geq \frac{(p-1)(p^2-8p+16)}{p^9} 
\left( \frac{p^3}{10} \right)^2 = \frac{1}{100} -  O\left(\frac{1}{p}\right)\,.
%\frac{\frac{1}{2}p(p^2-8p)}{100p^3}
%\geq \frac{p-8}{200p}\geq \frac{1}{400}
\]
Therefore, the success probability can be lower bounded by a constant for sufficiently large $p$. Furthermore, the
computations in the appendix show that we find the solutions of the
polynomial system~(\ref{EQQ}) by solving univariate polynomials of
degree six or less. This leads to an efficient quantum algorithm
because the roots of these polynomials can be computed with
a polylogarithmic number of operations.

\section{Conclusion and Outlook}
We have introduced the Hidden Polynomial Function Graph Problem as a
generalization of a particular abelian Hidden Subgroup Problem.  We
have shown that the standard approach for HSPs can be successfully
applied to this problem and leads to an efficient quantum algorithm
for quadratic and cubic polynomials over prime fields.  A
generalization of all the methods to non-prime fields $\mathbb{F}_d$
is straightforward.  The Fourier transform over $\Z_p$ has to be
replaced by the Fourier transform over $\mathbb{F}_d$ which can be
implemented efficiently \cite{DHI:03}.

The central points of interest for future research are the
generalization to polynomials over rings (admitting a Fourier
transform), polynomials of higher degree, multivariate polynomials,
and a broader class of functions.  Moreover, it would be important to
find real-life problems which could be reduced to our black-box
problem and the problems defined in \cite{Childs}.  \\

The authors acknowledge helpful discussions with D.~Bacon and
D.~Janzing. TD was supported under ARO/DTO quantum algorithms grant
number W911NSF-06-1-0379.

\appendix
\section{Analysis of the Cubic Case}
In this appendix we use Buchberger's algorithm\footnote{We use the 
lexicographical order of monomials.} to show that the 
ideal that is generated by the polynomials of Eq.~(\ref{EQQ}) 
contains the elements
\begin{eqnarray*}
b+xc+yd-u&=&0\\
c+g_1d^5+g_2d^4+g_3d^3+g_4d^2+g_5d+g_6&=&0 \\
d^6+h_1d^5+h_2d^4+h_3d^3+h_4d^2+h_5d+h_6&=&0
\end{eqnarray*}
for a subset of the tuples $(x,y,u,v,w)$ which we refer to as regular
cases. From these equations inequality~(\ref{ineq2}) follows
directly because there are at most six solutions for $d$
and each value of $d$ determines $b$ and $c$ uniquely. Additionally, we 
consider non-regular cases in order to establish the inequality for 
all $(x,y,u,v,w)$ with certain $x$ and $y$. In the latter cases we obtain
at most ten solutions since there are at most five possible values
for $d$ and for each of those values there are at most two pairs
$(b,c)$ which lead to a solution of the system.

\subsection{Buchberger's Algorithm in Regular Cases}\label{a1}
Before computing S-polynomials following Buchberger's algorithm we
reduce the polynomials of Eq.~(\ref{EQQ}) with the linear polynomial,
i.e., we eliminate $b$ in the second and third polynomial equation
with the substitution
\[
b=u-xc-yd\,.
\]
This leads to the equations
\begin{eqnarray}\label{a}
c^2+c_1cd+c_2c+c_3d^2+c_4d+c_5&=&0\\
c^3+d_1c^2d+d_2c^2+d_3cd^2+d_4cd+d_5c\label{b}
+d_6d^3+d_7d^2+d_8d+d_9&=&0
\end{eqnarray}
with the coefficients
\[
\begin{array}{lllllll}
c_1:=\frac{2y}{x+1}&\quad&
c_2:=\frac{-2 u}{x+1}&\quad&
c_3:=\frac{y(y+1)}{x(x+1)}&\quad&
c_4:=\frac{-2uy}{x(x+1)}\\
c_5:=\frac{u^2-v}{x(x+1)}&\quad&
d_1:=\frac{-3xy}{1-x^2}&\quad&
d_2:=\frac{3ux}{1-x^2}&\quad&
d_3:=\frac{-3y^2}{1-x^2}\\
d_4:=\frac{6uy}{1-x^2}&\quad&
d_5:=\frac{-3u^2}{1-x^2}&\quad&
d_6:=\frac{y(1-y^2)}{x(1-x^2)}&\quad&
d_7:=\frac{3uy^2}{x(1-x^2)}\\
d_8:=\frac{-3u^2y}{x(1-x^2)}&\quad&
d_9:=\frac{u^3-w}{x(1-x^2)}
\end{array}
\]
Here and in the remainder of this section we assume that all occurring
denominators are unequal to zero. We reduce Eq.~(\ref{b}) with
Eq.~(\ref{a}) and obtain the polynomial
\begin{equation}\label{c}
cd^2 + e_1cd + e_2 c + e_3 d^3 + e_4 d^2 +e_5 d +e_6
\end{equation}
where we have
\[
\begin{array}{lll}
e_1:=\frac{d_4-c_1d_2-c_2d_1-c_4+2c_1c_2}{d_3-c_1d_1-c_3+c_1^2}&\quad\quad&
e_2:=\frac{d_5-c_2d_2-c_5+c_2^2}{d_3-c_1d_1-c_3+c_1^2}\\
e_3:=\frac{d_6-c_3d_1+c_1c_3}{d_3-c_1d_1-c_3+c_1^2}&\quad\quad&
e_4:=\frac{d_7-c_3d_2-c_4d_1+c_1c_4+c_2c_3}{d_3-c_1d_1-c_3+c_1^2}\\
e_5:=\frac{d_8-c_4d_2-c_5d_1+c_1c_5+c_2c_4}{d_3-c_1d_1-c_3+c_1^2}&\quad\quad&
e_6:=\frac{d_9-c_5d_2+c_2c_5}{d_3-c_1d_1-c_3+c_1^2}
\end{array}
\]
After these reductions we compute the reduced S-polynomial
\begin{equation}\label{fpoly}
cd +f_1c+f_2d^4+f_3d^3+f_4d^2+f_5d+f_6
\end{equation}
of the polynomials~(\ref{a}) and~(\ref{c}).
We have the coefficients
\begin{eqnarray*}
f_1&:=&\frac{e_6-e_2e_4+e_1e_2e_3}{e_5-e_1e_4-e_2e_3+e_1^2e_3}\\
f_2&:=&-\frac{e_3^2-c_1e_3+c_3}{e_5-e_1e_4-e_2e_3+e_1^2e_3}\\
f_3&:=&-\frac{2e_3e_4-c_1e_4-e_1e_3^2-c_2e_3+c_3e_1+c_4}{e_5
-e_1e_4-e_2e_3+e_1^2e_3}\\
f_4&:=&-\frac{e_3e_5-c_1e_5+e_4^2-e_1e_3e_4-c_2e_4+c_3e_2+
c_4e_1+c_5}{e_5-e_1e_4-e_2e_3+e_1^2e_3}\\
f_5&:=&-\frac{e_3e_6-c_1e_6+e_4e_5-e_1e_3e_5-c_2e_5+c_4e_2
+c_5e_1}{e_5-e_1e_4-e_2e_3+e_1^2e_3}\\
f_6&:=&-\frac{e_4e_6-e_1e_3e_6-c_2e_6+c_5e_2}{e_5-e_1e_4
-e_2e_3+e_1^2e_3}
\end{eqnarray*}
The reduced S-polynomial of the polynomials~(\ref{c})
and~(\ref{fpoly}) is
\begin{equation}\label{g}
c+g_1d^5+g_2d^4+g_3d^3+g_4d^2+g_5d+g_6
\end{equation}
where we have
\[
\begin{array}{lllll}
g_1:=-\frac{f_2}{f_1^2-e_1f_1+e_2}&\quad&
g_2:=-\frac{f_3-f_1f_2+e_1f_2}{f_1^2-e_1f_1+e_2}&\quad&
g_3:=-\frac{f_4-f_1f_3+e_1f_3-e_3}{f_1^2-e_1f_1+e_2}\\
g_4:=-\frac{f_5-f_1f_4+e_1f_4-e_4}{f_1^2-e_1f_1+e_2}&&
g_5:=-\frac{f_6-f_1f_5+e_1f_5-e_5}{f_1^2-e_1f_1+e_2}&&
g_6:=\frac{f_1f_6-e_1f_6+e_6}{f_1^2-e_1f_1+e_2}
\end{array}
\]
The reduced S-polynomial of the polynomials in Eq.~(\ref{fpoly})
and~(\ref{g}) is
\begin{equation}\label{eqq}
d^6+h_1d^5+h_2d^4+h_3d^3+h_4d^2+h_5d+h_6 
\end{equation}
with the coefficients
\[
\begin{array}{llllll}
h_1:=\frac{g_2+f_1g_1}{g_1}&\quad&
h_2:=\frac{g_3+f_1g_2-f_2}{g_1}&\quad&
h_3:=\frac{g_4+f_1g_3-f_3}{g_1}\\
h_4:=\frac{g_5+f_1g_4-f_4}{g_1}&&
h_5:=\frac{g_6+f_1g_5-f_5}{g_1}&&
h_6:=\frac{f_1g_6-f_6}{g_1}
\end{array}
\]
After this step we stop Buchberger's algorithm since the
S-polynomials~(\ref{g}) and~(\ref{eqq}) are the polynomials we are
looking for. This computation is only possible for regular tuples
$(x,y,u,v,w)$, i.e., all denominators are non-vanishing.  

\subsection{Characterization of Regular Cases}  
In the preceding section, all denominators are unequal to zero if the
disequalities
\[
\begin{array}{lll}
0\not=x(x+1)&\quad
0\not=x(1-x^2)\\
0\not=d_3-c_1d_1-c_3+c_1^2&\quad
0\not=e_5-e_1e_4-e_2e_3+e_1^2e_3\\
0\not=f_1^2-e_1f_1+e_2&\quad
0\not=g_1
\end{array}
\]
are satisfied.  The substitution of $c_j$, $d_j$, $e_j$, $f_1$,
and $g_1$ with their expressions in $x,y,u,v,w$ leads to the
disequalities
\begin{eqnarray*}
0&\not=&x(x+1)\\
0&\not=&x(1-x^2)\\
0&\not=&y(y+x+1)\\
0&\not=&(y-x-1)(vy+vx+v-u^2)\\
0&\not=&(r_3v^3 + r_2 v^2 + r_1 v + r_0)(x+1) \\
0&\not=&(y+1)(y+x)
\end{eqnarray*}
where we have the coefficients
\begin{eqnarray*}
r_0&:=&w^2xy^3+w^2y^3+2w^2x^2y^2+4w^2xy^2
+2w^2y^2+w^2x^3y\\&&+3w^2x^2y+3w^2xy
+4u^3wxy+w^2y+4u^3wy+u^6\\
r_1&:=&-3u(y+x+1)(2wxy+2wy+u^3)\\
r_2&:=&3u^2(y^2+xy+y+x^2+2x+1)\\
r_3&:=&-(y-x-1)^2(y+x+1)
\end{eqnarray*}
For the following analysis we separate the factors of the disequalities
into two sets: The first set contains all factors which only
depend on $x$ and $y$ and the second set contains all factors
which also depend on $u$, $v$, or $w$.

\subsection{Analysis of Non-Regular Cases}
In this section we discuss the polynomial system for non-regular
tuples $(x,y,u,v,w)$, i.e., one or more of the denominators of
Sec.~\ref{a1} vanish.  We assume that all factors of the denominators
which solely depend on $x$ and $y$ are unequal to zero since the other
cases can be discarded in the analysis of Sec.~\ref{sec7}. The
remaining disequalities which depend on the $u$, $v$, and $w$ are
\[
0 \not = v(y+x+1)-u^2  \quad {\rm and} \quad 
0 \not = r_3v^3 + r_2 v^2 + r_1 v + r_0 \,.
\]

First, we assume that $v(y+x+1)-u^2 =0$. Then Buchberger's algorithm
leads to the polynomial system
\begin{eqnarray*}
b+xc+yd-u&=&0\\
cd^2 + e_1cd + e_2 c + e_3 d^3 + e_4 d^2 +e_5 d +e_6
&=&0\\
{\tilde f}_1 c + {\tilde f}_2 d^4 +{\tilde f}_3 d^3 +
{\tilde f}_4 d^2 + {\tilde f}_5 d + {\tilde f}_6&=&0
\end{eqnarray*}
where we have 
\[
{\tilde f}_j := (e_5-e_1e_4-e_2e_3+e_1^2e_3)f_j
\]
with the $f_j$ of Section~\ref{a1}. If ${\tilde f}_1 \not = 0$ then we
substitute $c$ in the second equation with 
\[
c=-\frac{1}{{\tilde f}_1}\left( 
{\tilde f}_2 d^4 +{\tilde f}_3 d^3 +{\tilde f}_4 d^2 +{\tilde f}_5 d 
+ {\tilde f}_6\right)\,.
\]
This substitution leads to a polynomial in $d$ which 
always has degree six since 
\[
{\tilde f}_2 = -\frac{(y+1)(y+x)}{(x-1)^2x}
\]
is always non-zero for the $x$ and $y$ we consider. Hence, there are
at most six solutions for $d$ and inequality~(\ref{ineq2}) also holds
in this non-regular case because $b$ and $c$ are uniquely defined by
the value of $d$. For ${\tilde f}_1 = 0$ we have the system
\begin{eqnarray*}
b+xc+yd-u&=&0\\
cd^2 + e_1cd + e_2 c + e_3 d^3 + e_4 d^2 +e_5 d +e_6
&=&0\\
{\tilde f}_2 d^4 +{\tilde f}_3 d^3 +
{\tilde f}_4 d^2 + {\tilde f}_5 d + {\tilde f}_6&=&0
\end{eqnarray*}
where again ${\tilde f}_2 \not = 0$ is always true. In this case there 
are at most four solutions for $d$. Furthermore, it follows from
Eq.~(\ref{a}) that for each value of $d$ there are at most two
solutions for $c$. Since $b$ is uniquely defined by $c$ and $d$ there
are at most eight solutions. 

For $v(y+x+1)-u^2 \not=0$ we consider the case $r_3v^3 + r_2 v^2 + r_1
v + r_0 =0$. We obtain the reduced S-polynomial
\begin{equation}\label{ppp}
{\tilde g}_1 d^5 + {\tilde g}_2 d^4 + {\tilde g}_3 d^3 + 
{\tilde g}_4 d^2 + {\tilde g}_5 d^1 + {\tilde g}_6 
\end{equation}
where we have the coefficients
\[
{\tilde g}_j:=(f_1^2-e_1f_1+e_2)g_j
\]
with the $g_j$ of Section~\ref{a1}.
There are always at most five solutions for $d$ since
\[
{\tilde g}_1=\frac{-y(y+1)(y+x)(y+x+1)^2}
{x(x-1)(y-x-1)(vy+vx+v-u^2)}
\]
shows that polynomial~(\ref{ppp}) is not the zero polynomial. Therefore,
the system has at most ten solutions as the discussion of the preceding 
case shows.


\begin{thebibliography}{}
\bibitem{BCD06} D. Bacon, A. Childs, and W. van Dam, {\em Optimal
measurements for the dihedral hidden subgroup problem}, Chicago
Journal of Theoretical Computer Science, Article 2, 2006.

\bibitem{BCD05} D. Bacon, A. Childs, and W. van Dam, {\em From optimal
measurements to efficient quantum algorithms for the hidden subgroup
problem over semidirect product groups}, Proc. of the 46th Symposium
on Foundations of Computer Science, 2005, pp. 469-478.

\bibitem{BD07} A. Childs and W. van Dam, {\em Quantum algorithm for a
generalized hidden shift problem}, Proc. 18th ACM-SIAM Symposium on
Discrete Algorithms, 2007, pp. 1225-1234.

\bibitem{BL95} R. Boneh and R. Lipton, {\em Quantum cryptanalysis of
hidden linear functions}, Proc. Advances in Cryptology, Lecture Notes
in Computer Science {\bf 963}, 1995, pp. 424--437.

\bibitem{Childs} A. Childs, L. Schulman, and U. Vazirani, {\em Quantum
algorithms for hidden nonlinear structures}, Personal communication
and talk given at the QIP Workshop 2007, Brisbane, Australia, January
30 -- February 3, 2007.

\bibitem{EH99} M. Ettinger and P. H\o yer, {\em A quantum observable for
the graph isomorphism problem}, quant-ph/9901029.

\bibitem{DHI:03}
W. van Dam, S. Hallgren, L. Ip,
{\em Quantum Algorithms for some Hidden Shift Problems},
SIAM Journal on Computing, Volume 36, Issue 3, pp. 763-778.

\bibitem{Gathen} J. von zur Gathen, J. Gerhard: Modern Computer
Algebra, Cambridge University Press, 2003.

\bibitem{Hallgren02} S. Hallgren, {\em Polynomial-time quantum
algorithms for Pell's equation and the principal ideal problem},
Proc. 34th ACM Symposium on Theory of Computing, 2002, pp. 653--658.

\bibitem{Hallgren05} S. Hallgren, {\em Fast quantum algorithms for
computing the unit group and class group of a number field},
Proc. 37th ACM Symposium on Theory of Computing, 2005, pp. 468--474.

\bibitem{Hallgren}
S. Hallgren, C. Moore, M. R\"otteler, A. Russell, and P. Sen,
{\em Limitations of quantum coset states for graph isomorphism},
Proc. of 38th ACM Symposium on Theory of Computing, 2006, pp. 604 -- 617.

\bibitem{HW06}
A. Harrow, A. Winter, {\em How many copies are needed for state discrimination?}, \texttt{http://arxiv.org/abs/quant-ph/0606131}, 2006.

\bibitem{HW94} P. Hausladen and W. K. Wootters, {\em A `pretty good'
measurement for distinguishing quantum states}, Journal of Modern
Optics {\bf 41}, no. 12, pp. 2385--2390.

\bibitem{ISS07} G. Ivanyos, L. Sanselme, and M. Santha, {\em Quantum
algorithm for the hidden subgroup problem in extraspecial groups},
Proc. of 24th Annual Symposium on Theoretical Aspects of Computer
Science, Lecture Notes in Computer Science {\bf 4393}, 2007,
pp. 586--597.

\bibitem{MR95}
R. Motwani and P. Raghavan, {\em Randomized algorithms}, Cambridge University Press, 1995.


\bibitem{Regev02} O. Regev, {\em Quantum computation and lattice
problems}, Proc. 43rd Symposium on Foundations of Computer Science,
2002, pp. 520--529.

\bibitem{SV05} A. Schmidt and U. Vollmer, {\em Polynomial time quantum
algorithm for the computation of the unit group of a number field},
Proc. 37th ACM Symposium on Theory of Computing, 2005, pp. 475--480.

\bibitem{Shor97} P. W. Shor, {\em Polynomial-time algorithms for prime
factorizations and discrete logarithms on a quantum computer}, SIAM
Journal on Computing {\bf 26}, 1997, pp. 1484--1509.
\end{thebibliography}
\end{document}